\documentclass[10pt,twocolumn,superscriptaddress,aps,prl,showpacs,preprintnumbers,amsmath,amssymb,floatfix,nofootinbib]{revtex4}

\usepackage[latin9]{inputenc}
\usepackage{color}

\usepackage{graphicx}
\usepackage{esint}
\usepackage[unicode=true,bookmarks=false,breaklinks=false,pdfborder={0 0 1},backref=false,colorlinks=true, citecolor=blue]{hyperref}

\newcommand{\pr}{\,{ \rm P}\, }

\makeatletter
\@ifundefined{textcolor}{}
{%
 \definecolor{BLACK}{gray}{0}
 \definecolor{WHITE}{gray}{1}
 \definecolor{RED}{rgb}{1,0,0}
 \definecolor{GREEN}{rgb}{0,1,0}
 \definecolor{BLUE}{rgb}{0,0,1}
 \definecolor{CYAN}{cmyk}{1,0,0,0}
 \definecolor{MAGENTA}{cmyk}{0,1,0,0}
 \definecolor{YELLOW}{cmyk}{0,0,1,0}
}

\usepackage{graphicx}
\usepackage[english]{babel}
\usepackage{amsmath}
\usepackage{amssymb}
\definecolor{orange}{rgb}{1,0.5,0}
\usepackage{bm}

\newcommand{\re}{\mbox{Re}}
\newcommand{\im}{\mbox{Im}}
\newcommand{\sgn}{\mbox{sgn}}

\begin{document}
\title{Flight of a heavy particle nonlinearly coupled to a quantum bath}
\author{Mohammad F. Maghrebi}
\email[Corresponding author: ]{magrebi@umd.edu}
\affiliation{Joint Quantum Institute, NIST/University of Maryland, College Park, Maryland 20742, USA}
\affiliation{Joint Center for Quantum Information and Computer Science, NIST/University of Maryland, College Park, Maryland 20742, USA}
\author{Matthias Kr\"{u}ger}
\affiliation{Max-Planck-Institut f\"{u}r Intelligente Systeme - Heisenbergstr. 3, D-70569 Stuttgart, Germany}
\affiliation{Institut f\"{u}r Theoretische Physik IV, Universit\"{a}t Stuttgart - Pfaffenwaldring 57, D-70569 Stuttgart, Germany}
\author{Mehran Kardar}
\affiliation{Department of Physics, Massachusetts Institute of Technology, Cambridge, Massachusetts 02139, USA}

\begin{abstract}
Fluctuation and dissipation are by-products of coupling to the `environment.'
The Caldeira-Leggett model, a successful paradigm of quantum Brownian motion, views the environment
as a collection of harmonic oscillators {\it linearly coupled} to the system.
However, symmetry considerations may forbid a linear coupling, e.g. for a neutral particle in quantum electrodynamics.
We argue that absence of linear couplings can lead to a fundamentally different behavior.
Specifically, we consider a heavy particle quadratically coupled  to quantum fluctuations of the bath.
In one dimension the particle undergoes anomalous diffusion, unfolding as a power-law distribution in space, reminiscent of L\'{e}vy flights.
We suggest condensed matter analogs where similar effects may arise.
\end{abstract}

\maketitle

\section{Introduction}
Brownian motion is the prototype of classical stochastic phenomena;
its quantum counterpart is more complex due to the intricacies of quantum mechanics.
Dissipation of energy, as well as stochastic fluctuations, are consequences of couplings of the system (e.g. a particle)
to a reservoir.
A paradigm of quantum Brownian motion is the Caldeira-Leggett model in which  the reservoir is modeled as a collection of harmonic oscillators,
an adequately realistic prescription at low temperatures~\cite{Caldeira81,Caldeira83-1}.
This approach has been applied to a plethora of systems such as the spin-boson problem~\cite{Leggett87}, quantum tunneling~\cite{Caldeira81,Caldeira83-2}, quantum interference~\cite{Caldeira85},
including several platforms in solid state physics~\cite{WeissBook}.
The notion of quantum Brownian motion has also been used in other contexts, see e.g. Ref.~\cite{Erdos12}.

A crucial hypothesis in the Caldeira-Leggett approach is the linearity of the interaction between system and reservoir~\cite{Caldeira81}.
The resulting model is sufficiently simple to
allow for integrating out the environmental coordinates. Much analytical progress is then possible, mostly with functional integral methods~\cite{Caldeira81, Caldeira83-1,Caldeira85,Ingold88,Hu92}.
Further justifying the above approach,
{\it additional nonlinearities} in the coupling to the environment, which can be treated perturbatively~\cite{Hu93},
can be argued to be irrelevant, in the sense
that the system is effectively describable by linear couplings under certain conditions
(see, for example, App.~C in Ref~\cite{Caldeira83-2}).
Further elaborations include an environment composed of interacting fermions~\cite{Guinea84} or spins~\cite{Stamp00,Stamp06};
the latter~\cite{Stamp06} demonstrating anomalous diffusion of a quantum particle coupled to a spin bath.
In general, an environment not described by harmonic oscillators linearly coupled to the system may give rise to qualitatively different behaviors beyond the prototypical Caldeira-Leggett model.

Here, we consider systems in which {\it linear couplings to the environment are forbidden by symmetry},
leading to fundamentally different behaviors due to allowed nonlinear interactions.
An important example is a  heavy (classical) {\it neutral} particle, where the absence of
charge results in quadratic coupling to the electromagnetic fields of the environment.
We find that such a particle, coupled quadratically to the bath,  undergoes anomalous diffusion
 at {\it low temperatures}, characterized by a power law distribution in space,
in sharp contrast to the Gaussian distribution of classical (or linearly coupled quantum) Brownian motion.
Confirming the inherently quantum nature of this result,
we find that usual (Gaussian) Brownian motion is restored
at  finite temperatures for sufficiently long times, albeit with a diffusion constant depending anomalously on temperature.

The manuscript is organized as follows. In Sec.~\ref{sec:EQ}, we introduce the general model under consideration, together with its equations of motion and action. In Sec.~\ref{Sec. Linear coupling}, we review previous results for different linear couplings, before starting the detailed analysis for a quadratic coupling in Sec.~\ref{Sec. Quadratic coupling}. While these discussions  are limited to zero temperature, the case of finite temperature is analyzed in Sec.~\ref{Sec. Finite temperature}. We conclude in Sec.~\ref{Sec. Discussion and outlook}.

\section{Stochastic motion}
\subsection{Equations of motion and action}\label{sec:EQ}
Consider a general framework for a classical particle of mass $m$ undergoing dissipative and stochastic motion,
 $X(t)$ denoting its position at time $t$.
 The equation of motion  in an external potential $V(X)$, cast in frequency domain, is
\begin{equation}
  -m \omega^2 X(\omega) + \mathcal{F}\{{\partial_X V}\}(\omega)= F[ X,\omega]\,, \label{eq:1}
\end{equation}
with $\mathcal{F}\{\cdot\}$ denoting the Fourier transform.
The as yet unspecified dissipative and fluctuating force $F$ is a function(al) of $\omega$ and $X$, arising from the microscopic degrees
of freedom in the environment.
Note that a classical description of the particle is valid only at times large compared to intrinsic quantum time scales, such as $\hbar/mc^2$ in the relativistic context considered later ($\hbar$ and $c$ denote Planck's constant and the speed of light, respectively).
The quantum character of the bath is captured with appropriate $F[ X,\omega]$,
which can be expanded as $F[X,\omega]= F_0(\omega)+\chi(\omega)X(\omega)+\cdots$, keeping the lowest order terms describing fluctuations and dissipation.
Here, $F_0(\omega)$ is the stochastic force acting on a pinned particle,
while $\chi(\omega)X(\omega)$ is a (linearized) deterministic force, which may renormalize the intertia $m$.

In the absence of the external potential, the equation of motion can be  organized to first order in $X$  as
\begin{equation}\label{Eq. stochastic eom}
  \left[-m\omega^2-\chi(\omega)\right]X(\omega)=F_0(\omega)\,.
\end{equation}
Near equilibrium, the fluctuation-dissipation theorem (FDT) relates the fluctuations of the stochastic force $F_0(\omega)$ to dissipation by
\(
\im \chi(\omega)=\hbar^{-1}\tanh(\beta\hbar \omega/2) C_F(\omega)
\),
with $\beta $ the inverse temperature, and $C_F(t)=(1/2) \left\langle F_0(t) F_0(0) +F_0(0) F_0(t)\right\rangle$ the force-force correlator~\cite{Kubo12}---for useful FDT relations, see Appendix~\ref{app:FDT}.
For  linear coupling to an environment consisting of harmonic oscillators, $F_0$ is Gaussian distributed.
Consequently, the distribution function of  $X$ is also Gaussian, its form entirely captured by the two-point correlation function $\left\langle X(t)X(0)\right\rangle$.
By  virtue of the FDT, this implies that the dispersion of a particle linearly coupled to the environment can be fully constructed from (the imaginary part of) $\chi(\omega)$.
In the absence of linear couplings to the bath, explicit computation of the $n$-point functions  (moments) of the coordinate $X$ indicates that the behavior
can be drastically different at low temperatures.

As a concrete example,  consider environment coordinates described by a scalar field $\Phi(t,x)$ in one dimension.
A convenient way to define the model is via the action
\begin{equation}\label{Eq. action for Phi}
  S=\frac{1}{2}\int \!\! dt\int \!\! dx \,\left[\frac{1}{c^2}\left(\partial_t \Phi\right)^2 -(\partial_x \Phi)^2\right]+S_I\, ,
\end{equation}
which is simply a continuum description of harmonic oscillators.
This expression arises in a relativistic quantum field theory with light speed $c$, and may also emerge as an effective description
in a condensed matter system, with $c$ describing the propagation of low-energy modes~\cite{Sachdev07}.
Furthermore, the particle at position $X$ is locally coupled to the field at (or near) $x=X$ through $S_I$.

\subsection{Linear coupling} \label{Sec. Linear coupling}

Let us take the interaction of the particle with the environment as
	\begin{equation}\label{Eq. L Phi}
		S_I=\int \!\!dt \, {L}_\Phi\!\left(t, X(t)\right)\, ,
	\end{equation}
	where ${L}_\Phi$ is local and linear in $\Phi$. In particular, we consider two cases: a) ${L}_\Phi \propto\Phi$, and b) ${ L}_\Phi\propto\partial_x \Phi$. The former describes, for example, the interaction of a (non-relativistic)  charged particle with the `quantum electrodynamic' (QED) field of the bath, while the latter describes a heavy impurity in a Luttinger liquid where $\partial_x\Phi$ represents the local density~\cite{Fisher96}.
The resulting force is $F_{\rm a}\sim \,\partial_x\Phi$, and $F_{\rm b}\sim \,\partial_x^2\Phi$, evaluated at the position of the particle.
To compute force fluctuations for the particle at rest, we set $X=0$. Since $F_0(t)\equiv F[X=0,t]$ is linear in $\Phi$, its fluctuations are Gaussian, and can be computed directly from Eq.~(\ref{Eq. action for Phi}). Dissipation can be deduced from fluctuations via the FDT as
	\(
	 \im \chi_{\rm a}(\omega)\sim  \omega
	  \)
	   and
	    \(
	     \im \chi_{\rm b}(\omega)\sim \omega^3
	      \)
	      for small $\omega${ as we have shown in Appendix \ref{App. Linear}}.
      The former produces a typical friction force, proportional to velocity $f_{\rm a}\propto - \partial_t X$;
      the latter giving rise to a friction force of Abraham-Lorentz form, $f_{\rm b}\sim \partial_t^3 X$, consistent with Ref.~\cite{Fazio13}.
      In both cases, the mean square displacement (MSD) of the particle at long times behaves as~\cite{Hakim85,WeissBook,Sinha92}
      \begin{equation}\label{Eq. logarithm}
	      \lim_{t\gg\tau} \left \langle (X(t)-X(0))^2 \right\rangle \sim \log \frac{t}{\tau}\;,
      \end{equation}
      where $\tau$ is a short time scale; see Appendix \ref{App. Linear} for a derivation.
The distribution $P(t,X)$ of the particle's position at time $t$,
starting from $X=0$, is a Gaussian with a variance that expands as $\log t$. While for a general form of $\chi(\omega)$ the variance is not always logarithmic~\cite{WeissBook}, a linear coupling always leads to a Gaussian distribution function in $X$.

\subsection{Quadratic coupling} \label{Sec. Quadratic coupling}
More generally, the coupling of a system to the bath can be expanded as a power series in $\Phi$,
generically  including the first order term. An important exception is when a linear term is forbidden on the basis of symmetry.
A specific example is provided by a {\it neutral} (but polarizable) particle in the bath corresponding to Eq.~(\ref{Eq. action for Phi});
this may represent the true vacuum with zero-point QED fluctuations,
or alternatively the ground state of a condensed matter system with an emergent gauge field.
The force acting on a charged particle is invariant under the simultaneous transformation $e \to -e$ and $\Phi \to -\Phi$ where $e$ is the charge, and must thus have an expansion of the form $F\sim ``e \Phi" +``\Phi^2"+ \cdots$ (derivatives are not shown).
For a neutral particle
there can be no linear term, and the interaction starts at quadratic order in $\Phi$.
Specifically, let us consider a perfectly reflecting particle such that $\Phi\left(t, X(t)\right)=0$, i.e., subject to Dirichlet boundary conditions, see also Ref.~\cite{Dalvit00}.
In one dimension, this condition cuts the space into two disconnected regions which can be treated independently.
The energy-momentum tensor for the scalar field can be computed from the Lagrangian density ${\cal L}=\frac{1}{2c^2}\left(\partial_t \Phi\right)^2 -\frac{1}{2}(\partial_x \Phi)^2$ as  $T^{\mu}_{\nu}=\frac{\partial {\cal L}}{\partial (\partial_\mu \Phi)}\partial_\nu \Phi-{\cal L}\delta^{\mu}_{ \nu}$, from which the force can be computed as $F=T^{1}_{1}(x=0^+)-T^{1}_{1}(x=0^-)$. With Dirichlet boundary conditions, this force takes the form
\begin{equation}\label{Eq. Force}
  F=\frac{1}{2}(\partial_x \Phi_L)^2-\frac{1}{2}(\partial_x \Phi_R)^2\;,
\end{equation}
(again evaluated at the position of the particle) where $L$ ($R$) denotes the field evaluated at the left (right) of the particle.
Indeed, the force starts at quadratic order in $\Phi$ in harmony with our symmetry argument\footnote{One can also formally define an action $S_I$ from which the force in Eq.~\eqref{Eq. Force} is obtained~\cite{Dean10}, for example, $S_I=\alpha\int dt  \Phi(t,X(t))^2$ with $\alpha\to\infty$.}.

\begin{figure}[t]
     \centering
     \includegraphics[width=9cm]{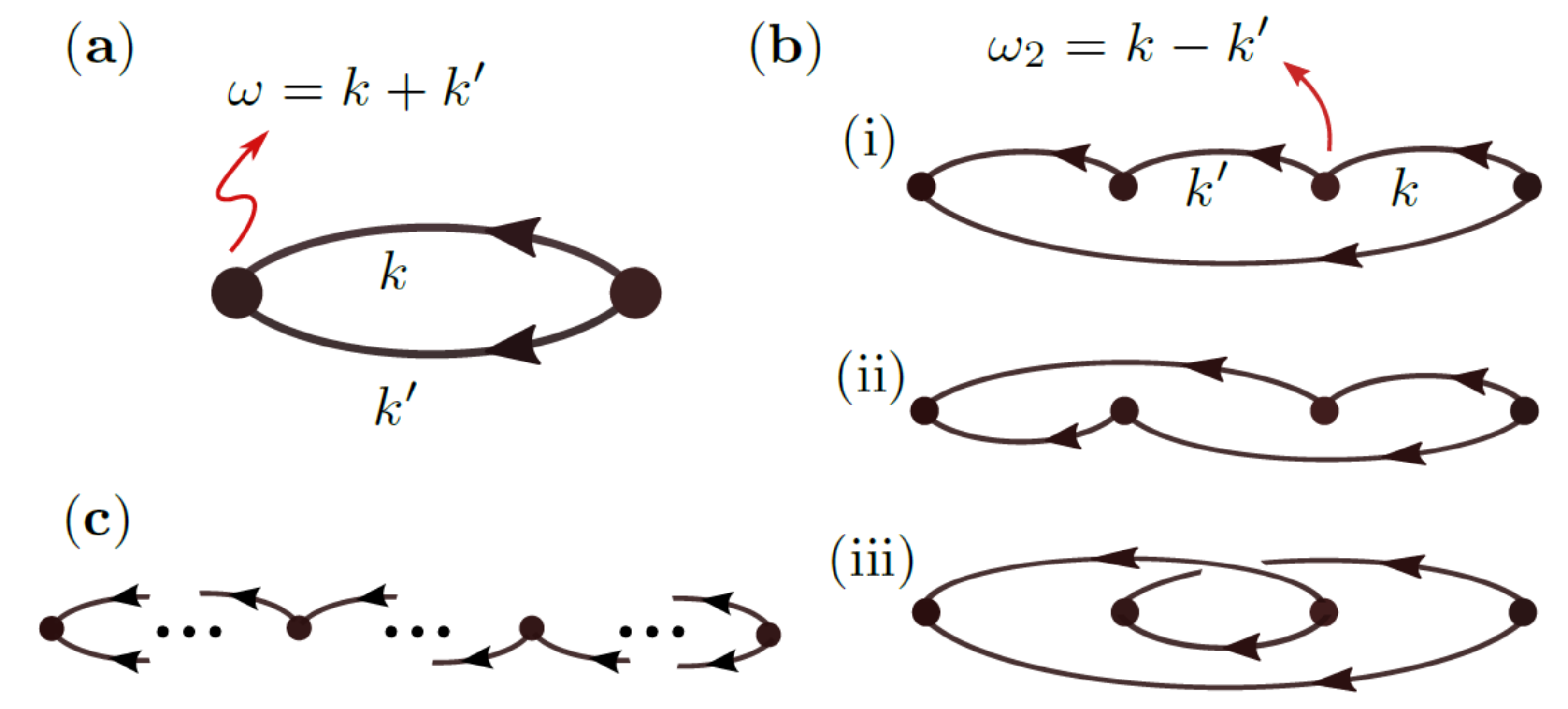}
      \caption{Diagrams contributing to the connected $n$-point correlation function in Eq.~(\ref{Eq. x to n}):  Each vertex (solid dot) is an insertion of $F_0$ with two incoming/outgoing lines representing photons.
     {\bf  (a)} Two-point function; two photons are created by the first insertion of $F_0$ and annihilated by the second. Energy conservation dictates $\omega=k+k'$.
     {\bf  (b)} Three distinct diagrams contribute to four-point functions. Diagrams (i) and (ii) have internal vertices with one incoming and one outgoing line. The corresponding frequency [for example, the second vertex of (i)]
      generates a pole when $k=k'$.
     {\bf  (c)} A typical $n$-point function with all---except the first and the last---vertices attached to both an incoming and an outgoing line.} \label{Fig}
   \end{figure}

Employing a second--quantized approach, the field (on either side, dropping the $L/R$ subscripts) is expanded as
\begin{equation}\label{Eq. Phi expanded}
  \Phi(t,x)=\int_0^{\infty}\!\!\frac{dk}{2\pi}\sqrt{{\frac{2}{k}}}\,\,  \sin kx \, \left(e^{-ikt} a_k+ e^{ikt} a_k^\dagger\right)\;,
  \end{equation}
  where $k$ is the momentum, and annihilation and creation operators satisfy
  \(
    [a_k, a_{k'}^\dagger]=2\pi \delta(k-k').
  \)
(We have set $\hbar=c=1$ for convenience.)
Note that $\Phi$ vanishes at $x=0$ where the particle is located, and the above algebra enforces the canonical commutation relation of the field and its conjugate momentum.
To compute force fluctuations, the following simple diagrammatic representation is useful:
Let us first consider $\langle 0|F_0(\omega) F_0(-\omega)|0\rangle$ where $|0\rangle$ represents vacuum, and $F_0(\omega)$ is the Fourier transform of Eq.~(\ref{Eq. Force}) with $X=0$ pinned.
$F_0(-\omega)$ is a bilinear operator in $\Phi$, and, upon acting on vacuum $|0\rangle$, creates two `photons' with momenta $k$ and $k'$ such that $\omega=k+k'$; these photons are then annihilated by $F_0(\omega)$~\cite{Neto93}, see Fig.~\ref{Fig}(a).
Each line in the figure is accompanied by a factor $k$ which is easily derived by noting the derivatives in Eq.~(\ref{Eq. Force})
and the normalization in Eq.~(\ref{Eq. Phi expanded}).
One then finds $\langle 0|F_0(\omega) F_0(-\omega)|0\rangle\sim \int_0^\omega dk k(\omega-k)\sim \omega^3$ for small positive $\omega$.  An explicit computation for a trajectory $X(t)$ yields the friction force \cite{Fulling76,Ford82,Jaekel92} (restoring $\hbar$ and $c$),
\begin{align}
\im \chi(\omega)=\frac{\hbar}{6\pi c^2}\omega^3, \,\text{ or }\, f=\frac{\hbar }{6\pi c^2}\dddot X\,.
\end{align}

Computing the second moment of displacement directly via Eq.~(\ref{Eq. stochastic eom}), or by following the same line of reasoning as in case (b) of Sec.~\ref{Sec. Linear coupling}, we find a similar result for the MSD that depends logarithmically on time,
\begin{equation}\label{Eq. log t}
\langle (X(t)-X(0))^2\rangle = \frac{\lambda^2}{3\pi^2}\log\frac{t}{\tau}\,.
\end{equation}
$\lambda\equiv \hbar/mc$ is the de Broglie wavelength, and $\tau=\hbar/mc^2$.
(We have thus verified that the independently calculated $\im \chi$ and $C_F$ satisfy the FDT.)
The force $f$ can be interpreted as the friction of vacuum~\cite{Golestanian99}, and is related to the dynamic Casimir effect.
Since our model is Lorentz invariant, $f$ can only depend on derivatives of velocity.
The two-point fluctuations are thus almost identical to the linear model (b),
and indeed the MSD takes a form similar to Eq.~(\ref{Eq. logarithm}).
Nevertheless, as we shall see, higher-point fluctuations drastically change such a correspondence.

Higher moments of the particle's position can be  computed through Eq.~(\ref{Eq. stochastic eom}); the $n$-th cumulant
is given by
  \begin{align}\label{Eq. x to n}
    {\big\langle} (X(t)-X(0))^n{\big\rangle}_c &=\int \prod_{i=1}^n \frac{d\omega_i}{2\pi}\, \left(e^{-i \omega_i t}-1\right)  R({\omega_i})\nonumber \\ \times \,{\big\langle } 0{\big|}& F_0({\omega_n})\cdots F_0({\omega_1}){\big |}0{\big\rangle}_c\;.
  \end{align}
  The subscript $c$ denotes the connected component of the correlation functions\footnote{For example, ${\langle} (X(t)-X(0))^4{\rangle}_c\equiv{\langle} (X(t)-X(0))^4{\rangle}-3 {\langle} (X(t)-X(0))^2{\rangle}^2$.}, quantifying deviations from Gaussian behavior~\cite{Risken84};
$R$ is the response function obtained as the inverse of the bracket in Eq.~(\ref{Eq. stochastic eom}),
  \begin{equation}\label{Eq. Response fn}
    R(\omega)=\frac{1}{ -m \omega^2-\chi(\omega)-i \epsilon \,\sgn(\omega)}\,,
  \end{equation}
with $i \epsilon$ inserted to ensure causality.
After collecting inertial terms from $\re\chi$ (redefining $m$ as the total mass), higher powers of frequency in real and imaginary parts of $\chi$ ($\im\chi\sim~\omega^3$) will not be important compared to the leading $\omega^2$ term at long times.
Hence, somewhat surprisingly, $\chi(\omega)$ can be dropped from Eq.~(\ref{Eq. Response fn})!

  To go beyond the MSD, we first compute force fluctuations with $n=4$ (contributions from $L$ and $R$ cancel out for odd $n$).
  There are three distinct diagrams contributing to $n=4$ as depicted in Fig.~\ref{Fig}(b).
  In contrast to the two-point diagram (a), the first two diagrams (i, ii) in Fig.~\ref{Fig}(b) include vertices with an incoming and an outgoing line; the corresponding frequency is now the difference of the incoming and outgoing momenta,
 for example, $\omega_2=k-k'$ in  diagram (i).
 Thus there is a relatively large phase space, $k=k'$, for this frequency to vanish.
 This is not the case for the leftmost (rightmost) vertex as both incoming (outgoing) momenta must vanish in order to have $\omega_1=0$ ($\omega_4=0$).
  The significance of this observation is that the integrand in Eq.~(\ref{Eq. x to n}) is quite sensitive to the frequency poles in the response functions of Eq.~(\ref{Eq. Response fn}). In the vicinity of one such pole,
  the frequency dependence in the first line of Eq.~(\ref{Eq. x to n}) can be cast as
  \begin{equation}\nonumber
    \frac{\sin(\omega_i t/2)}{\omega_i}\frac{1}{\omega_i +i \epsilon }\,.
  \end{equation}
  Unless there are powers of $\omega_i$ coming from $\langle 0|F_0 \cdots F_0|0\rangle_c$ (which is not the case for  vertices with one incoming and one outgoing line, see Appendix \ref{App. Quadratic}), this expression is divergent near $\omega_i=0$. Therefore, we can use the identity $1/(\omega_i +i\epsilon)={\rm P}\frac{1}{\omega_i}-i\pi \delta(\omega_i)$ to evaluate the integral. In particular,  substituting the delta function in the last equation yields a linear dependence on $t$ since  $\lim_{\omega_i\to 0}\frac{\sin(\omega_i t/2)}{\omega_i}\sim t$.  Both diagrams (i) and (ii) in Fig.~\ref{Fig}(b) have two such vertices, and thus contribute to the fourth cumulant of the particle's displacement as
  \begin{equation}\label{Eq. x to 4}
    {\big\langle} (X(t)-X(0))^4{\big\rangle}_c^{(\rm i), (\rm ii)} \sim \lambda^2(v t)^2\; ,
  \end{equation}
where we have introduced a velocity scale $v$.
In contrast to the diagrams (i) and (ii), the contribution of  diagram (iii) grows as a power of $\log t$.
To obtain finite results from integrations, we  needed to introduce a cutoff frequency, $\Gamma$, beyond which the particle is assumed to be completely transparent. The velocity $v$ is related to this cutoff by $v/c=2\Gamma/m\ll 1$; the inequality following from the reasonable assumption that $\Gamma\ll m$.
Equation~(\ref{Eq. x to 4}) implies a rapid increase with time of the fourth moment, compared to the slow logarithmic growth of the MSD.
In striking contrast to a purely Gaussian process, where ${\big\langle} (X(t)-X(0))^n{\big\rangle}_c=0$  for $n>2$,
the fourth cumulant grows faster than the variance at long times,
 rendering the process more and more non-Gaussian as time evolves.

Given the above, we have computed all cumulants of the particle's displacement to the leading order at long times.
  Such computation requires the sum over all diagrams with any (even) number of vertices in which each vertex, except the first and the last one, is attached to one incoming and one outgoing line, see Fig.~\ref{Fig}(c).
  The  result for the sum of all such diagrams (with $\tilde\lambda=\lambda/\sqrt{6}\pi$) is
  \begin{equation}\label{Eq. x to n power law}
    {\big\langle} (X(t)-X(0))^n{\big\rangle}_c =\frac{{2\tilde \lambda}^2}{n-2} \, (v t)^{n-2}\,,
  \end{equation}
 to leading order in $t$, and for all (even) $n>2$ (see Appendix \ref{App. Quadratic} for details).
 All cumulants ($n>2$) grow as a power-law in $t$; the exponent $n-2$ coming from the number of vertices with both incoming and outgoing lines.
 Again, this is in sharp contrast to traditional Brownian motion where the cumulants for $n\geq3$ are subdominant to the variance.
 As a consistency check, we remark that the moments $M_n\equiv {\big\langle} (X(t)-X(0))^n{\big\rangle}$
satisfy the constraint $M_{n+m}^2\le M_{n}\times M_{n+2m}$ that simply follows from positivity of probabilities (Pawula theorem~\cite{Risken84}).

The long-time dominance of higher cumulants in Eq.~\eqref{Eq. x to n power law} indicates that the probability
distribution function is broad; indeed an infinity of experiments would be needed to measure finite moments.
Any experimental observations will be sampled from the distribution function $P(t,X)$ for the position of the particle at
time $t$, starting from $X=0$ at $t=0$. Such distribution
is consistent with the simple form
  \begin{equation}\label{Eq. Distribution}
    P(t,X)=
      \frac{{\tilde \lambda}^2}{|X|^3}\, , \qquad  \mbox{for}  \quad \tilde \lambda <|X|<vt\,.
  \end{equation}
  The region $|X|\lesssim\tilde \lambda$ requires a quantum mechanical treatment of the particle; for $|X|> vt \gg \tilde \lambda$, the distribution function $P$ is possibly nonzero, but suppressed at long times.
 Remarkably, the distribution function in Eq.~(\ref{Eq. Distribution}) is a power law, in contrast to the Gaussian distribution of the usual Brownian motion (or even quantum dispersion with linear coupling). The absence of linear coupling to the environment thus qualitatively changes  quantum Brownian motion. A notable feature of Eq.~(\ref{Eq. Distribution}) is the emergence of a wavefront that propagates with the speed $v$.
The probability distribution is peaked around $|X|\sim \tilde\lambda$, and its
only time dependence resides in the linear propagation of the wavefront in the tails.
Nevertheless, probability is conserved, $\int dX\, P(t,X)\approx 1$, since for $vt\gg \tilde\lambda$
the contribution from the tail is negligible.

Power-law tails of the distribution are reminiscent of L\'evy flights.
However, unlike the latter our process is not Markovian.
In particular, Eq.~(\ref{Eq. Distribution}) does not satisfy the Chapman-Kolmogorov equation that marks Markovian processes~\cite{Risken84}.
This reflects the long time correlations (memory or coherence) of the quantum bath at zero temperature. Introducing a finite temperature indeed makes the system Markovian at sufficiently long times.

\subsection{Finite temperature} \label{Sec. Finite temperature}
All the above models  acquire a finite mobility at finite temperature. This may be expected since thermal photons  colliding with the particle mimic a more classical Brownian motion. At a temperature $T$, the loss of coherence occurs over a time scale $t\sim \hbar/ k_B T$, leading to a Markovian process with the standard diffusion law: The MSD increases linearly with $t$, $\left\langle (X(t)-X(0))^2\right\rangle= 2 D(T) \, t$ with a finite, temperature-dependent, diffusion constant $D(T)$ that is related to mobility $\mu(T)$ via the Einstein relation $D(T)=k_B T \mu(T)$.
The distribution  becomes more and more Gaussian over time, with cumulants that are all linear in $t$, and thus subdominant at long times\footnote{This is because, for example, $\left\langle (X(t)-X(0))^4\right\rangle_c \sim t \ll \left\langle  (X(t)-X(0))^2\right\rangle_c^2 \sim t^2$ at long times.}.
Nevertheless, we stress that the anomalous diffusion in the absence of linear coupling [Eq.~(\ref{Eq. Distribution})] persists even at finite temperature for times $t\lesssim \hbar /k_B T$.

    Here, we consider in some detail the model in Section \ref{Sec. Quadratic coupling} at a finite temperature $T$.
    It is useful to first compute the two-point field correlator. Defining $\Phi'(t)\equiv\partial_x \Phi(t,0)$), we find (in units that $k_B=\hbar=c=1$)
    \begin{align}\label{Eq: two-point}
      \left\langle \Phi'(t) \Phi'(0) \right\rangle
      &\sim \int_0^\infty d k\, k \left[e^{-ikt} (n(k)+1)+e^{ikt} n(k)\right]\nonumber \\
      &\sim T^2 {\rm csch}^2(\pi t T),
    \end{align}
    where $n(k)=\left[\exp(k/T)-1\right]^{-1}$ is the Bose-Einstein factor. In the zero-temperature limit,
      \(
      \lim_{T \to 0}\left\langle \Phi'(t) \Phi'(0) \right\rangle\sim 1/t^2\,,
    \)
     indicating long-time memory, and thus non-Markovianity.
     The force-force correlator at finite temperature follows from Eq.~(\ref{Eq: two-point}) as
     \begin{align}\label{Eq. Finite T}
                \langle F_0(t) F_0(0)\rangle_c\sim T^4 {\rm csch}^4(\pi t T).
     \end{align}
    Equations (\ref{Eq: two-point}) and (\ref{Eq. Finite T}) show that the correlation functions are exponentially suppressed at long times.
    Let us recall the definition $C_F(t)\equiv \langle F_0(t) F_0(0)+F_0(0) F_0(t)\rangle/2$. We are interested in the long-time asymptotics, and thus expand the Fourier transform of Eq.~(\ref{Eq. Finite T}) at small frequencies to obtain
    \(
      C_{F}(\omega)\sim T^3 +{\cal O}\left(\omega^2\right)
    \).
   The response function is related to fluctuations via the FDT. Specifically, at high temperatures, $\im \chi\sim (\omega/T)C_F(\omega)$, hence, we obtain
   \begin{equation}\label{Eq: friction}
     \im \chi(\omega)\sim T^2\, \omega,
   \end{equation}
   leading to a friction force linear in velocity.
   It then follows that the mobility scales as
  \begin{equation}
    \mu(T) \sim T^{-2}\,,
  \end{equation}
   giving rise to an anomalous diffusion constant
     \(
     D(T)\sim 1/T\,
     \) as $T\to 0$.
  The divergence of the mobility as $T\to 0$ is expected since the model is Lorentz invariant in the absence of a background `photon' bath. Including $n$-point functions beyond $n=2$ does not change the above conclusions as shown in Appendix \ref{App. Finite T}.

\section{Discussion and outlook}\label{Sec. Discussion and outlook}
Fluctuation and dissipation in a quantum bath are intriguingly different from their classical counterparts.
We have shown that non-linear coupling to environment coordinates,
robustly dictated by symmetry considerations in certain settings,
can lead to yet another distinction:
A non-Gaussian power-law distribution (in one dimension) akin to a L\'{e}vy flight
(albeit non-Markovian).
Further studies in  higher dimensions should be of interest from both fundamental and practical perspectives; we briefly remark that the argument that led to the power-law dependence on time, namely the frequency poles in Eq.~(\ref{Eq. x to n}), appears to hold in all dimensions.

In this work, we have focused on a particle interacting with a scalar field in vacuum; however, similar considerations apply to a wide range of problems. First we remark that the Lorentz symmetry can be emergent (as in Luttinger liquid), while the `particle' may be an impurity in a condensed matter system. The key ingredient is a general symmetry principle that forbids a linear coupling to the environment.
As another example, consider the transverse-field Ising model at its critical point,
described by $\Phi^4$ theory~\cite{Sachdev07}. For a non-magnetic impurity that  couples symmetrically to $\uparrow$ and $\downarrow$ spins, the force should be symmetric under $\Phi \to -\Phi$, and start at  quadratic order. Neglecting the $\Phi^4$ term, such impurity undergoes the same power-law diffusion described above, assuming it obeys Eq.~(\ref{eq:1}). The $\Phi^4$ interaction, however, may change this behavior in subtle ways with different universal properties,
making for an interesting topic of future study.
\\

{\bf Acknowledgements:} We acknowledge useful discussions with S. Shadkhoo and C. Henkel. MM was supported by NSF PFC at JQI, NSF PIF, ARO, ARL, AFOSR, and AFOSR MURI.
  MaKr was supported by DFG grant No. KR 3844/2-1; MK by the NSF through grant No.
DMR-12-06323.

\appendix

\section{Fluctuation-dissipation theorem}\label{app:FDT}
In this appendix, we provide some useful forms of the FDT.
The response function $R$ defined by
  \begin{equation}\label{Eq. X from F}
    \langle X(t)\rangle -\langle X(0)\rangle =\int R(t-t') F_0(t')\,,
  \end{equation}
  is related to the correlation function, $C(t)=(1/2) \langle X(t) X(0) +X(0) X(t)\rangle$, via the FDT as~\cite{Kubo12}
  \begin{equation}\label{Eq. Im R from C finite T}
    \im R(\omega)=\hbar^{-1}\tanh(\beta \hbar \omega/2) C(\omega).
  \end{equation}
  At zero temperature, this equation takes the form
  \begin{equation}\label{Eq. Im R from C}
    \im R(\omega)=\hbar^{-1}\sgn(\omega) C(\omega).
  \end{equation}
  By computing the inverse Fourier transforms, one finds at zero temperature \cite{Sinha92}
  \begin{align}\label{Eq: MSD real time}
    &\frac{1}{2} \left\langle \left(X(t)-X(0)\right)^2\right\rangle \nonumber \\
    &=\frac{\hbar }{2\pi} \int_0^\infty dt' R(t')\left[\frac{2}{ t'}-\frac{1}{t'+t}-\pr \frac{1}{t'-t}\right],
  \end{align}
  where $\pr$ denotes the Cauchy principle value. In a typical situation, the response $R(t)$ defined in Eq.~\eqref{Eq. X from F}  approaches a nonzero constant value for times large compared to a relaxation time $\tau$, i.e., $R(t)=\mu$ for $t\gg\tau$, which upon the substitution in Eq.~\eqref{Eq: MSD real time} yields the result given in Eq.~(\ref{Eq. logarithm}).

  We can interchange the role of the dynamical variable and the force in Eq.~(\ref{Eq. X from F}). The equivalents of Eqs.~(\ref{Eq. Im R from C finite T}) and (\ref{Eq. Im R from C}) are given by
  \begin{equation}\label{Eq. FDT for D finite T}
    \im \chi(\omega)=\hbar^{-1}\tanh(\beta \hbar \omega/2) C_F(\omega),
  \end{equation}
  at finite temperature, and
  \begin{equation}\label{Eq. FDT for D}
    \im \chi(\omega)=\hbar^{-1}\sgn(\omega) C_F(\omega)\,,
  \end{equation}
  at zero temperature.

\section{Linear models: Dissipation and displacement}\label{App. Linear}
In this appendix, we present a derivation of the results obtained in Sec.~\ref{Sec. Linear coupling}.
Consider the quantum field defined by the action in Eq.~(\ref{Eq. action for Phi}). The field in free space can be written in a second-quantized basis as ($\hbar=c=1$)
\begin{equation}
  \Phi(t,x)=\int_{-\infty}^\infty \frac{dk}{2\pi} \frac{1}{\sqrt{2\omega_k}}\left(e^{i kx-i\omega_k t}a_k+e^{-i kx+i\omega_k t}a^\dagger_k\right),
\end{equation}
where $\omega_k=|k|$.

For a particle weakly coupled to the field $\Phi$ via Eq.~(\ref{Eq. L Phi}) with $L_\Phi\propto \Phi$, the force is $F_0(t)\sim \partial_x \Phi(t,0)$. Some algebra yields
\begin{equation}
  F_0(t)\sim i\int_0^\infty\frac{dk}{2\pi} \sqrt{\frac{k}{2}}\left[e^{-ikt} (a_k-a_{-k})-e^{ikt}(a^\dagger_k-a^\dagger_{-k})\right].
\end{equation}
In particular, for $\omega>0$, $F_0(\omega)=F_0^\dagger(-\omega)\sim i \sqrt{\omega}\,(a_\omega-a_{-\omega})$ where the subscript on $a$ denotes momentum. One can then easily see that $C_F(\omega)\sim \omega$; we have used the fact that $\langle 0|a_k a_{k'}^\dagger|0\rangle=2\pi \delta(k-k')$, while all other two-point correlators vanish. Equation (\ref{Eq. FDT for D}) then dictates $\im \chi(\omega)\sim \omega$. The response function in real time is given by $R(t)\sim 1$ for $t\gg\tau$, with some $\tau>0$.

On the other hand, if the particle is coupled to the field via $L_\Phi\propto \partial_x \Phi$, the force becomes $F_0(t)\sim \partial_x^2 \Phi(t,0)$. Then,
\begin{equation}
  F_0(t)\sim \int_0^\infty\frac{dk}{2\pi} {\frac{k^{3/2}}{\sqrt{2}}}\left[e^{-ikt} (a_k+a_{-k})+e^{ikt}(a^\dagger_k+a^\dagger_{-k})\right],
\end{equation}
and $F_0(\omega)=F_0^\dagger(-\omega)\sim \omega^{3/2}\,(a_\omega+a_{-\omega})$ for $\omega>0$. Hence, $C_F(\omega)\sim \omega^{3}$, which in turn implies, via Eq.~(\ref{Eq. FDT for D}), that $\im \chi(\omega)\sim \omega^3$. The response function (neglecting runaway solutions) becomes $R(t)\sim t$ for $t\gg\tau$ with some $\tau>0$. The linear term gives a ballistic propagation corresponding to  the second time derivative in the equation of motion.

Both response functions computed above yield a logarithmic growth of the MSD when plugged in Eq.~(\ref{Eq: MSD real time}).

\section{Nonlinear model: All moments of displacement}\label{App. Quadratic}
In this appendix, we compute higher moments of the distribution function defined in Sec.~\ref{Sec. Quadratic coupling}.
We consider the region $x>0$ with the corresponding force $\frac{1}{2}(\partial_x\Phi(t,0))^2$ (up to an unimportant negative sign); the contribution of the region $x<0$ gives rise to an overall factor of 2. It is useful to explicitly write $\Phi'(t)\equiv\partial_x \Phi(t,0)$ starting from Eq.~(\ref{Eq. Phi expanded}),
\begin{equation}
  \Phi'(t)=\int_0^\infty \frac{d k}{2\pi} \sqrt{2k} \, \left(e^{-ikt} a_k+e^{ikt} a^\dagger_k\right).
\end{equation}
An insertion of the field produces two incoming/outgoing lines, each of which picks up a factor of $\sqrt{2k}$ with $k$ the corresponding momentum. Each line is shared by two vertices thereby contributing a factor of $2k$. Combined with $1/2$ in the definition of the force, we can assign each line a factor of $k$. However, each vertex
gives an additional factor of 2 due to the choice of contracting the lines attached to it with the incoming/outgoing lines.
We also write the exponential factors in Eq.~(\ref{Eq. x to n}) in the form of sine functions, thus getting an additional factor of $2i$ per vertex.

These considerations yield
\begin{widetext}
\begin{equation}\label{Eq. Fn}
  \left \langle (X(t)-X(0))^n\right\rangle =2\times\frac{2^{n}(2i)^n}{m^n}  \int \left[\prod_{i=1}^n\frac{d\omega_i}{2\pi}\frac{\sin(\omega_i t/2)}{\omega_i^2+i\epsilon \sgn(\omega_i)} \right]\, {\cal F}_n\left(\{\omega_i\}\right) 2\pi\delta\!\left(\omega_1+\cdots+\omega_n\right),
\end{equation}
\end{widetext}
where the overall factor of 2 takes into account the contribution from both left and right half of the space, the delta function makes the energy conservation explicit, and ${\cal F}_n$ should be computed by summing all $n$-point force correlation functions.
Following the discussion in Sec. \ref{Sec. Quadratic coupling}, we take $\omega_{i}\approx 0$ for $2\le i \le n-1$, and $\omega_n\approx-\omega_1$; this allows us to treat $\omega_1\gg 1/t$ as a fast variable and the rest of frequencies as slow variables $\omega_i\lesssim 1/t$. The first and the last sine functions in Eq.~(\ref{Eq. Fn}) can be then combined, by averaging over the fast variable $\omega_1$, as
\begin{equation*}
  \sin(\omega_1 t/2)\sin(\omega_n t/2) \to -\frac{1}{2}\cos(\bar \omega t/2),
\end{equation*}
where we used conservation of energy, and defined $\bar \omega=\omega_2+\cdots+\omega_{n-1}$. Furthermore, the function ${\cal F}_n$ can be approximated as  ${\cal F}_n(\omega_1,0,\cdots,0, -\omega_1)\equiv {\cal F}_n(\omega_1)$. To compute the diagrams in the form of Fig.~1(c), notice that there are only two momenta running within each diagram: $k$ and $\omega_1-k$; this is due to the assumption that $\omega_i\approx 0$ for all vertices except the first and the last ones. We should sum over all such diagrams with each (except the first and the last) vertex connecting to lines with the same, but either of the two, momenta. It is then straightforward to show that
\begin{equation}
  {\cal F}_n(\omega_1)=\int_0^{\omega_1} \frac{dk}{2\pi} k (\omega_1-k)\left[k+ (\omega_1-k)\right]^{n-2} =\frac{\omega_1^{n+1}}{12\pi},
\end{equation}
where the bracket has simply organized the sum over all possible diagrams (an expansion of the bracket generates all such diagrams).
The integration over $\omega_1$ now gives
\begin{equation}
  \int_0^{\Gamma} \frac{d\omega_1}{2\pi}\frac{1}{\omega_1^4}\,{\cal F}_n(\omega_1)=\frac{\Gamma^{n-2}}{24\pi^2(n-2)},
\end{equation}
where the denominator is due to the response functions (due to the first and the last vertices), and $\Gamma(\ll m)$ is a frequency cutoff on the idealized Dirichlet boundary conditions, that is, the point particle is transparent for frequencies $\omega\gtrsim \Gamma$. Finally, the integral over all other frequencies together with $\cos(\bar \omega t/2)$ is computed as
\begin{align}
\int &\left[\prod_{i=2}^{n-1}\frac{d\omega_i}{2\pi}\frac{\sin(\omega_i t/2)}{\omega_i^2+i\epsilon \sgn(\omega_i)} \right] \cos\left[(\omega_2+\cdots+\omega_{n-1})t/2\right]\notag\\&= -\frac{i^n t^{n-2}}{2^{n-1}}.
\end{align}
In computing this integral, we have expanded the trigonometric function $\cos(\bar \omega t/2)$, and used
\begin{align}
  \begin{split}
  &\int_{-\infty}^{\infty} \frac{d\omega}{2\pi}\frac{\sin(\omega t/2)\cos(\omega t/2)}{\omega^2+i\epsilon\sgn(\omega)}=-\frac{it}{4},\\ &\int_{-\infty}^{\infty} \frac{d\omega}{2\pi}\frac{\sin^2(\omega t/2)}{\omega^2}=\frac{t}{4}.
  \end{split}
\end{align}

Putting all these together, we obtain Eq.~(\ref{Eq. x to n power law}).

\section{Finite temperature}\label{App. Finite T}

   In this appendix, we extend the results obtained in Sec.~\ref{Sec. Finite temperature} at finite temperature to higher-$n$-point functions using an approximate, simplified approach.
   At long times ($t\gtrsim \hbar/ k_B T$), we approximately have ($k_B=\hbar=c=1$)
   \begin{equation}
     a_k\approx a_k^\dagger \approx \sqrt{ n(k)}\approx \sqrt{\frac{T}{ k}}\,,
   \end{equation}
   and, Eq.~(\ref{Eq: two-point}) becomes
   \begin{equation}\label{Eq: PhiPrime}
      \langle \Phi'(t) \Phi'(0) \rangle \sim T \int \frac{d k}{2\pi} \cos (kt)=T \delta(t).
   \end{equation}
   At long times, the dynamics is dominated by the frictional term computed in Eq.~(\ref{Eq: friction}),
   \begin{equation}\label{Eq: Langevin equation}
     T^2 \dot X= F(t).
   \end{equation}
   It follows from Eq.~(\ref{Eq: PhiPrime}) that
   \begin{equation} \label{Eq: force cor fn}
     \langle F(t) F(t')\rangle_c \sim  T^2\delta(0) \delta(t-t') \sim T^3 \delta(t-t'),
   \end{equation}
   where $\delta(0)$ is defined by an integral over frequencies less than $T$, hence it is proportional to $T$. Combining Eqs.~(\ref{Eq: Langevin equation}) and (\ref{Eq: force cor fn}), we find that
   \begin{equation}
     \langle (X(t)-X(0))^2\rangle \sim \frac{1}{T^4} \int^t\int^t dt' dt''\langle F(t') F(t'')\rangle_c  \sim \frac{1}{T} \, t,
   \end{equation}
   consistent with the fact that $D(T)\sim1/T$. In order to compute the fourth moment, for example, first note that
   \begin{align}
   &\langle F(t) F(t') F(t'')F(t''')\rangle_c \nonumber \\
   &\sim T^5 \delta(t-t')\delta(t'-t'')\delta(t''-t''')\,.
   \end{align}
   It then follows that
   \begin{equation}
     \langle (X(t)-X(0))^4\rangle_c \sim \frac{1}{T^3}t\,.
   \end{equation}
  Similarly, we find that {\it all} connected moments of $X$ grow linearly with time at long times.

\end{document}